\documentclass{pasj01}
\begin{document}
\Received{}%{yyyy/mm/dd}
\Accepted{}%{yyyy/mm/dd}

%\title{Soft Diffuse Sources and Low-Temperature Plasma  in the Galactic Center Region}
\title{Origin of the low-temperature plasma in the Galactic center X-ray emission}

\author{
Shigeo \textsc{Yamauchi}\altaffilmark{1 ${\ast}$},
Miku \textsc{Shimizu}\altaffilmark{1}, 
Masayoshi \textsc{Nobukawa}\altaffilmark{2},
Kumiko K. \textsc{Nobukawa}\altaffilmark{1},
Hideki \textsc{Uchiyama}\altaffilmark{3},
and 
Katsuji \textsc{Koyama}\altaffilmark{4}
}

\altaffiltext{1}{Department of Physics, Nara Women's University, Kitauoyanishimachi, Nara 630-8506}
\email{yamauchi@cc.nara-wu.ac.jp}
\altaffiltext{2}{Faculty of Education, Nara University of Education, Takabatake-cho, Nara 630-8528}
\altaffiltext{3}{Faculty of Education, Shizuoka University, 836 Ohya, Suruga-ku, Shizuoka 422-8529}
\altaffiltext{4}{Department of Physics, Graduate School of Science, Kyoto University, \\
Kitashirakawa-oiwake-cho, Sakyo-ku, Kyoto 606-8502}

 %Do NOT move this preamble from here!

\maketitle

\begin{abstract}
The Galactic Center X-ray Emission (GCXE) is composed of high temperature ($\sim$7 keV) and low 
temperature ($\sim$1 keV) plasmas (HTP and LTP, respectively). 
The global structure of the HTP is roughly uniform over the Galactic center (GC)
region, and the origin of the HTP has been extensively studied. 
On the other hand, the LTP is more clumpy, and the origin has not been studied in detail.  
In the S XV He$\alpha$ line map, a pair of horn-like soft diffuse sources are seen at the symmetric positions 
with respect to Sagittarius A$^\star$. 
The X-ray spectra of the pair are well represented by an absorbed thin thermal plasma model 
of a temperature and $N_{\rm H}$ of 0.6--0.7 keV and  4$\times$10$^{22}$ cm$^{-2}$, respectively. 
The $N_{\rm H}$ values indicate that the pair are located near at the GC.  
Then the dynamical time scales of the pair are $\sim$10$^{5}$ yr. The Si and S abundances and the surface brightnesses 
in the the S XV He$\alpha$ line band are 0.7--1.2 and 0.6--1.3 solar, and  (2.0--2.4)$\times$10$^{-15}$ 
erg s$^{-1}$ cm$^{-2}$ arcmin$^{-2}$, respectively.
The temperature, abundances, and surface brightness are similar to those of the LTP in the GCXE, while
the abundances are far larger than those of known point sources, typically coronal active stars and RS CVn-type active binaries. 
Based on these results, possible origin of the LTP is discussed.
\end{abstract}

\section{Introduction}

The spectrum of the Galactic Diffuse X-ray Emission (GDXE) has strong K-shell transition lines of highly ionized atoms and neutral iron.  
The strongest are the He-like iron (Fe XXV He$\alpha$) and sulfur (S XV He$\alpha$) lines, which indicates 
that the GDXE is composed of a high-temperature plasma of 
$\sim$7 keV (HTP) represented by the Fe XXV He$\alpha$ line, and a low-temperature plasma of $\sim$1 keV (LTP)  
represented by the S XV He$\alpha$ line \citep{Uchiyama2013}. The other component is a power-law with the Fe I K$\alpha$ line.
The equivalent widths and scale heights of the Fe XXV He$\alpha$  and Fe I K$\alpha$ lines are position dependent in the
GDXE, and hence the GDXE is spatially and spectrally separated into the Galactic Center X-ray Emission (GCXE), 
the Galactic Ridge X-ray Emission (GRXE), and the Galactic Bulge X-ray Emission (GBXE) \citep{Yamauchi2016, Nobukawa2016, Koyama2018}.

A long standing question of the GDXE is its origin, whether it is integrated emission of point sources, diffuse plasma, or else. 
Using the deep Chandra observation in the 6.5--7.1 keV band, \citet{Revnivtsev2009} and \citet{Hong2012} made an X-ray Luminosity Function 
(XLF: the integrated flux of point sources as a function of the point source flux) down to the luminosity of $\sim4\times10^{29}$ erg~ s$^{-1}$, 
and resolved more than 80 \% flux of the GBXE into point sources. 
However, the profiles of the XLF and integrated spectra of the point sources were largely different between these authors, 
which led different prediction of the point source composition in the GBXE: RS-CVn type active binaries (ABs) 
and cataclysmic variables (CVs) with the mixing ratio of $\sim$2:1 (\cite{Revnivtsev2009}), or CVs dominant (\cite{Hong2012}).  
A possibility of these mismatch in  the mixing ratio of point sources would be that the authors ignore
the energy band difference of the compositions; they simply referred the XLF results in the 6.5--7.1 keV band (HTP), not the 
full energy band of the HTP and LTP.
Using the spectrum difference of  ABs, CVs, and GBXE in the 5--10 keV band, \citet{Nobukawa2016} suggested 
that the GBXE is composed of ABs and CVs with the mixing ratio of  $\sim$3:7.  
In the full energy band (1--10 keV), the compositions would not be only these point sources (ABs and CVs), 
but may include true diffuse plasma or even unknown objects. 
Therefore, the simple point source origin should be carefully re-examined for the HTP and LTP in a proper mixture of these two components.

This paper utilizes the GCXE for the study of the LTP origin, because more reliable spectra would be available due to 
the surface brightness of  $\sim$10 times larger than the GBXE and GRXE. 
We reports properties of a pair of soft diffuse sources NE and NW in the LTP map at the northeast and northwest of the Galactic center (GC). 
The diffuse sources have been noted by \citet{Wang2002} (Chandra) and \citet{Ponti2015} (XMM-Newton), but the detailed information has not 
been reported.
Based on the improved spectral and spatial information of Suzaku, the origin of NE and NW, 
and possible interpretation of the origin of the LTP in the GCXE are discussed. 
Throughout this paper, the distance to the GC is 8 kpc (e.g., \cite{Reid1993,Gillessen2009}), and quoted errors are in the 90\% confidence limits. 

\section{Observations}

Survey observations in the GC region were carried out with the X-ray Imaging Spectrometer (XIS; \cite{Koyama2007a}) 
onboard Suzaku \citep{Mitsuda2007}. This paper utilized these Suzaku data in the archive.
The XISs were composed of 4 CCD cameras placed on the focal planes of the thin foil X-ray Telescopes (XRT; \cite{Serlemitsos2007}). 
XIS\,1 was a back-side illuminated (BI) CCD, while XIS\,0, 2, and 3 were front-side illuminated (FI) CCDs. 
The field of view (FOV) of the XIS was \timeform{17'.8}$\times$\timeform{17'.8}.
 The data from the three sensors (XIS\,0, 1, and 3) were used for most of the observations, because XIS\,2 stopped working in 2006 November. 
Since the spectral resolution of the XIS was degraded due to the radiation of cosmic particles,
the spaced-row charge injection (SCI) technique was applied to restore the XIS performance \citep{Uchiyama2009}.
After removing hot and flickering pixels, we used the data of the ASCA grade 0, 2, 3, 4, and 6.
%%%

\section{Analysis and results}

The XIS pulse-height data are converted to Pulse Invariant (PI) channels using the {\tt xispi} software in the HEAsoft 6.19 
and the calibration database version 2016-06-07.
The data in the South Atlantic Anomaly, during the earth occultation, and at the low elevation angle 
from the earth rim of $<5^{\circ}$ (night earth) and $<20^{\circ}$ (day earth) are excluded.
Figure 1 shows the results of the Suzaku GC survey, covering  the full area of the GC by multiple pointings. 
The non X-ray background (NXB), estimated using {\tt xisnxbgen} \citep{Tawa2008}, is subtracted. 
To highlight the contrast between the HTP and LTP distributions, X-ray images of the Fe XXV He$\alpha$ (6.55--6.8 keV) and  
S XV He$\alpha$  (2.3--2.6 keV)  bands are separately made.
A pair of soft diffuse sources (NE and NW) are detected in two fields of the XIS.
The observation logs of the two fields, which include the pair sources NE and NW, and background (BGD), are given in table 1.
%%%%%%
\begin{table*}[t] %table 1
\begin{center}
\caption{List of data used for spectral analyses.}
\begin{tabular}{lcccc} \hline  %\\ [-6pt]
Observation ID  & Pointing Position                   & Observation time (UT)& Exposure time & Region$^{\ast}$  \\
 & ($l$, $b$) & Start -- End & (ks) & \\
\hline %\\ [-6pt]
%501009010& (\timeform{359.D9261}, \timeform{+0.D1779})	& 2006-09-29 21:26:07 --	2006-10-01 06:55:19	& 51.2    & BGD2\\
503007010& (\timeform{0.D3285}, \timeform{+0.D1690})     	& 2008-09-02 10:15:27 -- 2008-09-03 22:52:24	& 52.2    & NE\\
503072010& (\timeform{359.D5753}, \timeform{+0.D1669}) 	& 2009-03-06 02:39:12 -- 2009-03-09 02:55:25	& 140.6  & NW, BGD\\
\hline 
\end{tabular}
\end{center}
%\vspace{-10pt}
$^{\ast}$ See figure 1.\\
\end{table*}

%%%%%%%%%%% Figure 1 %%%%%%%%%%
\begin{figure*}
  \begin{center}
        \includegraphics[width=16cm]{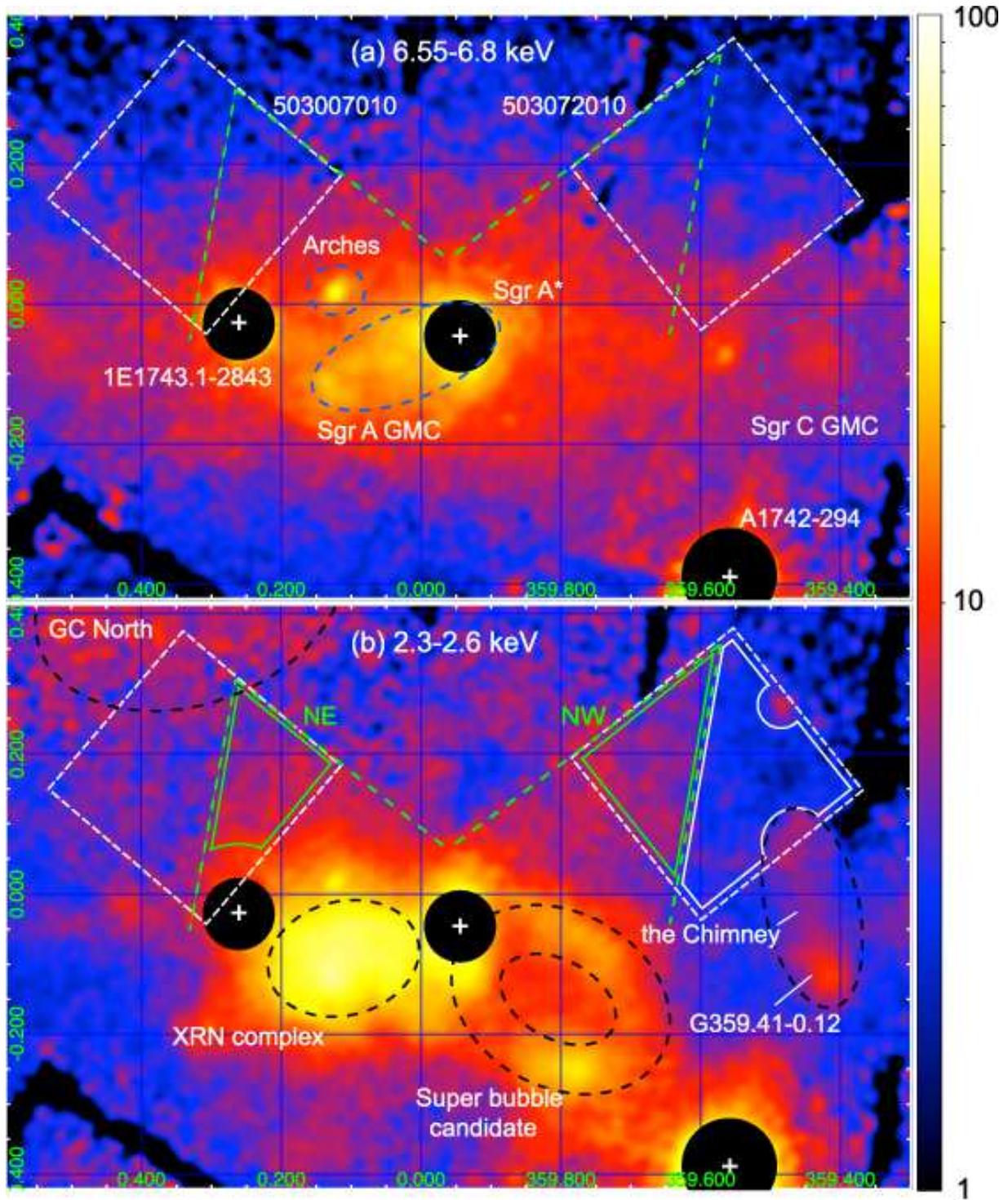}
  \end{center}
\caption{XIS images in the (a) 6.55--6.8 (Fe XXV He$\alpha$) and (b) 2.3--2.6 keV (S XV He$\alpha$) bands
in the Galactic coordinate. The color bar shows surface brightness in the logarithmic scale. 
The unit is  
%The number 100 corresponds to 
2$\times$10$^{-13}$ erg\,s$^{-1}$\,cm$^{-2}$\,arcmin$^{-2}$ for (a) and 
1$\times$10$^{-13}$ erg\,s$^{-1}$\,cm$^{-2}$\,arcmin$^{-2}$ for (b).
Bright point sources are masked by the black circles. A stray-light region of the brightest source A1742$-$294 is given by the large black circle.
The white dashed squares, the solid lines and the green horn-like lines indicate the XIS FOVs, the background region (BGD), and a pair of
soft diffuse sources (NE and NW), respectively.
The black dashed lines in (b) outline other LTP clumps, while the blue dashed lines in (a) are the regions of HTP clumps. 
In order to figure out the difference between the HTP and LTP structure, the XIS FOVs and horn-like structures in the LTP are also 
shown in the HTP image. 
}\label{fig:sample}
\end{figure*}

\subsection{Overview of soft diffuse sources in the GCXE}  %3.1

In the Fe XXV He$\alpha$ line map (HTP distribution), some slightly enhancement are found  
near at the giant molecular cloud complex (GMC) of Sagittarius (Sgr) A, Sgr C, and Arches cluster (blue dashed lines in figure 1a). 
The  Fe XXV He$\alpha$ line enhancement is about $\sim$10\% of the GCXE level. 
Thus global distribution of the  HTP is smooth in the full area of the GC.  
 In the LTP distribution, on the other hand, the S XV He$\alpha$ image shows  
a largely extended X-ray emission near at ($l$, $b$)$\sim$(\timeform{0.D3}, \timeform{0.D4}) (GC North, \cite{Nakashima2014}). 
The spectrum of this soft diffuse source is in collisional ionization equilibrium (CIE) with a temperature of 0.81 keV and solar abundances. 
Another soft diffuse  source is a super bubble candidate (G359.79$-$0.26 and G359.77$-$0.09) found by \citet{Mori2008}, 
\citet{Mori2009}, and \citet{Heard2013}, 
whose spectra are explained by an absorbed thermal plasma model in CIE  with temperatures of $\sim$ 1.0 and $\sim$ 0.7 keV, and
abundances of 1.1--1.7 and 1.0--1.4 solar, respectively.
A notable soft diffuse source  is  found around the Sgr A X-ray reflection nebula (XRN) near Sgr A$^{\star}$ (hear, XRN complex), 
firstly reported by \citet{Park2004}.  
This source has different morphology in the Fe XXV He$\alpha$ image, slim and faint, which corresponds  to the GMC complex 
(named,  Sgr A GMC), or Sgr A XRN.  For this source, however, no spectral information has been available.
Other soft diffuse sources are found near  the Sgr C GMC (the Chimney and G359.41-0.12), firstly reported by \citet{Tsuru2009}.  
The temperatures are $\sim$1.2 and $\sim$0.9 keV with abundances of $\sim$1.7 solar.
In addition to these soft diffuse sources, we can see a pair of horn-like diffuse sources  NE and NW, 
at the north (in the Galactic coordinate) of the GC. 

\subsection{X-ray spectra of soft diffuse sources, NE and NW} %3.2

\begin{figure*}[t] %Figure 2
  \begin{center}
       \includegraphics[width=8cm]{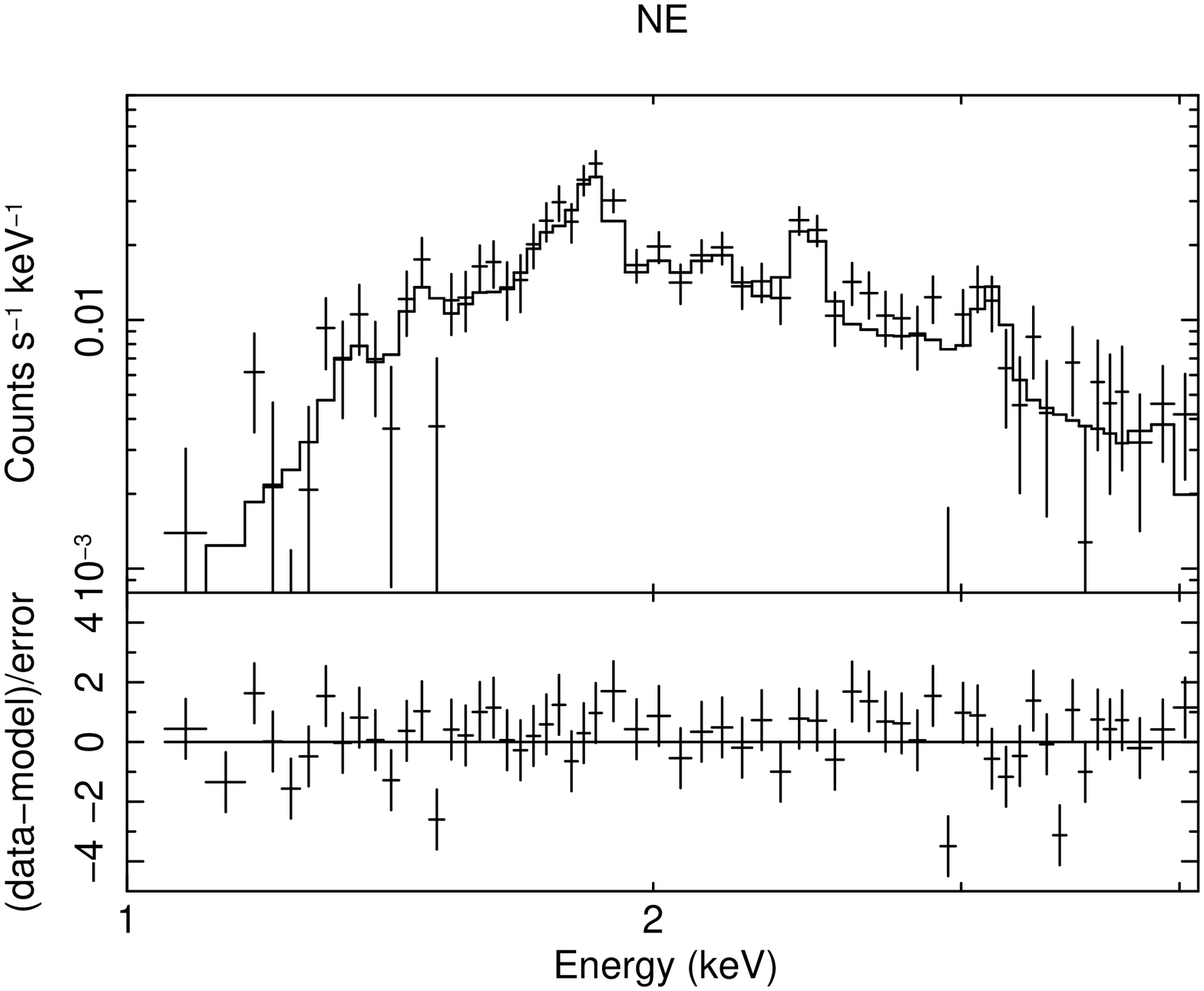}
       \includegraphics[width=8cm]{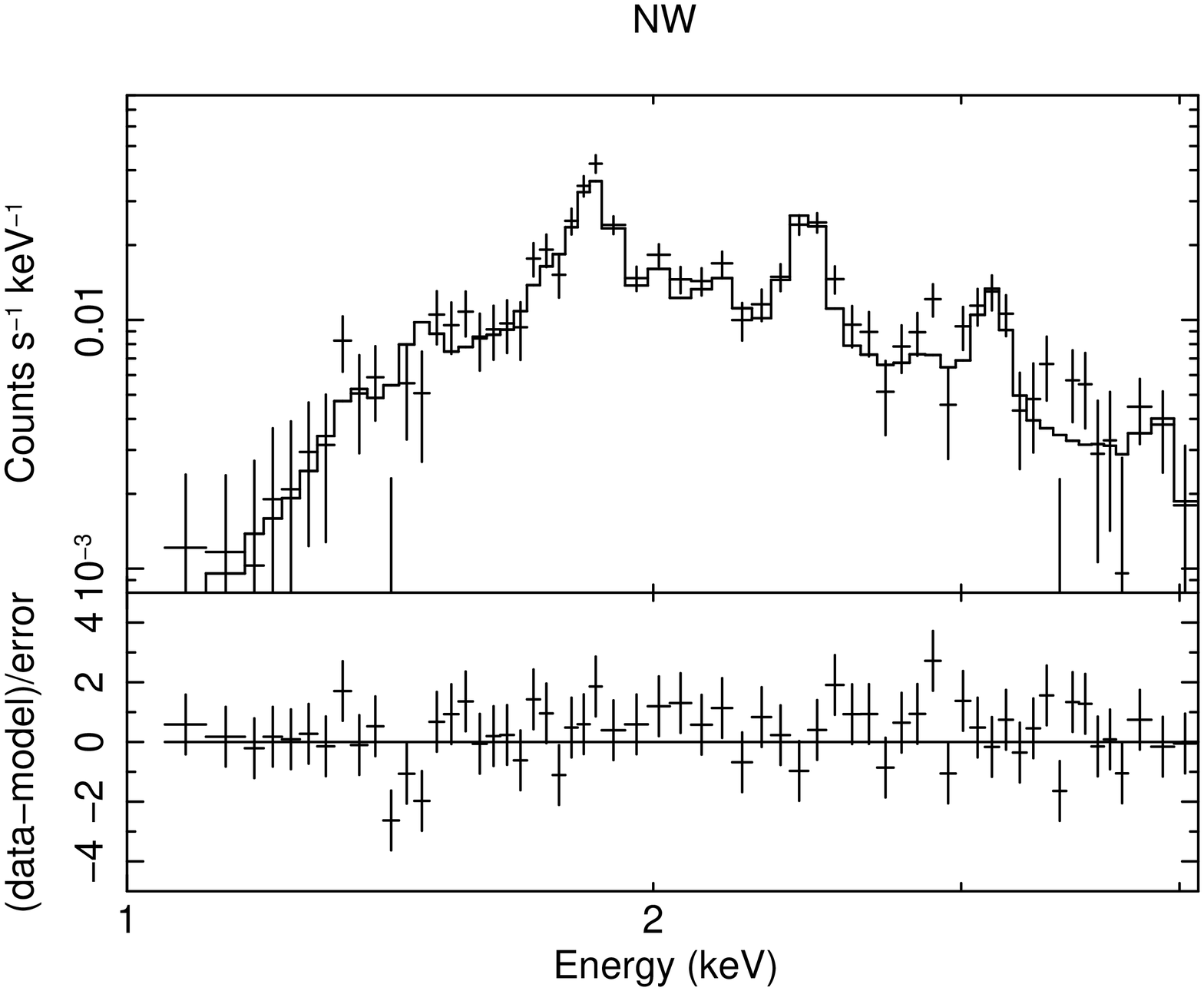}
   \end{center}
  \caption{X-ray spectra of NE and NW (upper panels) and the residuals from the best-fit model (lower panels).
Only the spectra obtained with the FI sensors in the 1--4 keV band are displayed for brevity. 
}\label{fig:sample}
\end{figure*}

The X-ray spectra of NE and NW after the subtraction of the NXB  (see section 2) are made from the source regions. 
The BGD region  is selected from a nearby blank sky  of similar Galactic latitude to those of NE and NW. % ($\Delta b$ = 0).  
The X-ray spectrum of BGD is made with the same process as NE and NW.  
In order to make the spectra of NE, NW, and BGD with good statistics, 
we utilize the two XIS fields, which include a large fraction  of NE, NW, and  BGD with long exposure times (table 1). 
The flux of the NXB are $\sim$9\% and $\sim$14\% of the total counts of the  background region (BGD) 
in the 1--8 keV band for FI and BI, respectively, and the statistical error is less then 1\%. 
Thus, the uncertainty caused by the NXB subtraction is not significant in the following spectral analysis. 
The flux of the BGD spectrum is fine-tuned, taking account of  the longitude and latitude differences 
between BGD and the pair sources (NE and NW) and 
using the e-folding longitude and latitude scales of the GCXE of \timeform{0.D63} and \timeform{0.D26}, respectively
\citep{Uchiyama2013,Yamauchi2016, Koyama2018}. 
The fine-tuning factor is 1.3 for NE and 1.1 for NW. 
Then, the BGD spectrum multiplied by this fine-tuning factor is subtracted from the NE and NW spectra.  
The resultant  NE and NW spectra are shown in figure 2, in which many emission line features are found.  
In order to increase photon statistics, the spectra with the FI sensors (XIS 0 and 3)  are co-added, 
but the XIS 1 spectrum is treated separately, because the response functions of the FIs and BI are different.
Response files, Redistribution Matrix Files (RMFs) and Ancillary Response Files (ARFs) are made using
{\tt xisrmfgen} and {\tt xissimarfgen} \citep{Ishisaki2007}, respectively. 
The abundance tables,  and  the atomic data of the lines and continua of the thin thermal plasma are taken from  
\citet{Anders1989} and  ATOMDB 3.0.9, respectively.

The NE and NW spectra are fitted with a CIE plasma model of solar abundances, {\tt vapec} in XSPEC version 12.9.0u. 
This model is rejected ($\chi^2$/d.o.f. of 163/119 and 177/119, respectively) with residuals at the Si, S, and Ar lines.  
Therefore,  the NE and NW spectra are re-fitted by the same  CIE model, 
but the abundances of Si, S, and Ar (=Ca) are treated as free parameters. 
Then an improved fit is obtained with $\chi^2$/d.o.f. of 147/116 and 158/116, 
respectively\footnote{Exactly speaking, this model is not accepted from the statistical point of view, 
which may be due to systematic errors caused in the BGD subtraction process. 
Taking account of the possible systematic errors, we regard the model is a good  approximation of the NE and NW spectra.}.
The best-fit model is shown in figure 2, while the best-fit parameters are given in table 2.
The flux in the S XV He$\alpha$ line band (2.3--2.6 keV) are  
2.4$\times$10$^{-15}$ erg s$^{-1}$ cm$^{-2}$ arcmin$^{-2}$ for NE and 
2.0$\times$10$^{-15}$ erg s$^{-1}$ cm$^{-2}$ arcmin$^{-2}$ for NW.

\begin{table}
\caption{The best-fit parameters of NE and NW.}\label{tab:first}
\begin{center}   
 \begin{tabular}{lcc}
      \hline
      Parameter 	& \multicolumn{2}{c}{Value}  \\
                            & NE & NW\\
      \hline
        $N_{\rm H}$ (cm$^{-2}$) 		&  (4.4$^{+0.3}_{-0.4}$)$\times$10$^{22}$  	& (4.2$^{+0.6}_{-0.2}$)$\times$10$^{22}$\\
        $kT_{\rm e}$ (keV)			& 0.64$^{+0.11}_{-0.07}$ 					&  0.71$^{+0.06}_{-0.12}$  \\
        Si$^{\ast}$ (Solar)			& 0.7$^{+0.3}_{-0.2}$				 	&  1.2$\pm$0.3  \\
        S$^{\ast}$ (Solar)			& 0.6$^{+0.3}_{-0.2}$ 					&  1.3$\pm$0.3  \\
        Ar=Ca$^{\ast}$ (Solar)		& 2.3$\pm$1.1		 					&  3.3$^{+1.2}_{-1.1}$  \\
        Others$^{\ast}$ (Solar)		& 1 (fixed) 							&  1 (fixed)   \\
        Normalization$^{\dag}$ 		&  (4.0$^{+2.2}_{-1.5}$)$\times$10$^{-4}$		&   (1.7$^{+1.6}_{-0.4}$)$\times$10$^{-4}$   \\ \hline
	$\chi^2$/d.o.f.  				& 147/116  							&  158/116   \\
   \hline
    \end{tabular}
\end{center}
 $^{\ast}$ Abundance relative to the solar value \citep{Anders1989}.\\
$^{\dag}$ Defined as 
10$^{-14}$$\times$$\int n_{\rm H} n_{\rm e} dV$ / 4$\pi D^2\Omega$ (cm$^{-5}$ arcmin$^{-2}$),
where $n_{\rm H}$ , $n_{\rm e}$, $D$ and $\Omega$ are   hydrogen density (cm$^{-3}$), 
electron density (cm$^{-3}$),  distance (cm) and  solid angle (arcmin$^2$) of the source, respectively.\\
 \end{table}
%%%%%

\section{Discussion}

 Chandra found many point sources in the GC region (e.g., \cite{Muno2009}).
The flux of the integrated point sources in the regions of NE and NW is  only
$\sim$10\%, and hence the contribution of the resolved point sources for NE and NW would be negligible.
The surface brightness of NE and NW is nearly the same level  of the nearby GCXE.  
The $N_{\rm H}$ values of the pair sources, NE and NW, are $\sim$4$\times$10$^{22}$ cm$^{-2}$ (table 2), 
roughly consistent with those of the point sources located at the GC region 
\citep{Sakano2000,Sakano2002}. Therefore, here and after, we assume that NE and NW are  located near the GC region at the distance of 8 kpc.  

The pair of soft diffuse sources, NE and NW,  have horn-like structures standing above the Galactic plane (see, figure 1b).
Assuming a cone shape geometry with a diameter of the base of $\sim$15$'$ (35 pc) and a height of $\sim$15$'$,
the volume ($V$) is estimated to be $V$ = 3.3$\times$10$^{59}$ cm$^{3}$. 
Using the best-fit volume emission measure, and assuming a filling factor = 1 and $n_{\rm e}$ = 1.2$n_{\rm H}$, 
where $n_{\rm e}$ and $n_{\rm H}$ are the electron and hydrogen densities, respectively, 
we obtain the mean hydrogen density ($n_{\rm H}$), the thermal energy ($E_{\rm th}$),  gas mass ($M_{\rm gas}$), 
and sound velocity ($c_{\rm s}$) for NE to be $n_{\rm H}$=0.3 cm$^{-3}$, $E_{\rm th}$=3$\times$10$^{50}$ erg, 
$M_{\rm gas}$=120$M_{\odot}$, and $c_{\rm s}$=4$\times$10$^7$ cm s$^{-1}$.
For NW, we obtained $n_{\rm H}$=0.2 cm$^{-3}$, $E_{\rm th}$=2$\times$10$^{50}$ erg,
$M_{\rm gas}$=80$M_{\odot}$, and $c_{\rm s}$=4$\times$10$^7$ cm s$^{-1}$.
The dynamical time scale ($t_{\rm dyn}$) is estimated to be $t_{\rm dyn}$$\simeq$1$\times$10$^5$ yr. 

Most of these physical  parameters of NE and NW are consistent with those of middle-aged SNRs. 
However, the plasma size is larger than typical middle-aged SNRs, and the horn-like morphology 
is largely different from that of a single SNR. 
The physical parameters and the dynamical time scales of  NE and NW are similar to each other, 
and the pair positions are symmetric with respect to Sgr A$^\star$. These suggest that the pair, NE and NW, 
originated from the same event, possibly a past activity in the GC region. 

Remarkable features in the LTP map are the presence of many bright soft diffuse sources 
\citep{Mori2008,Mori2009,Tsuru2009,Heard2013,Nakashima2014,Ponti2015}, 
including NE and NW, near the GC.
These soft diffuse sources have similar temperature and Si--S abundances to those of the LTP in the GCXE \citep{Uchiyama2013}.   
In the scenario of the point source origin for the LTP, a candidate source in the  luminosity range of $>$10$^{30}$ erg s$^{-1}$ 
has been regarded to be ABs with a thermal spectrum of temperature $\gtrsim$1 keV \citep{Sazonov2006,Warwick2014}. 
However, the fraction of the resolved point sources of the GCXE in this luminosity range is less than a few 10\% 
\citep{Muno2003, Revnivtsev2007}. 
Therefore, the major contribution of point sources should come from the lowest luminosity range of 
$\sim$10$^{28}$--10$^{30}$ erg s$^{-1}$.  In this luminosity range, the candidate source may  not be only ABs, but includes 
coronal active stars (CAs) with the temperature of  $\lesssim$1 keV (e.g., \cite{Gudel2004, Pandey2008}). 
In this case, the number of required point sources is more than $\sim$10$^5$--10$^6$, because the total LTP luminosity of the 
GCXE is $\sim$10$^{36}$ erg s$^{-1}$ \citep{Uchiyama2013}.   
Since this huge number of point sources would lead a uniform LTP distribution, the presence of many bright  soft diffuse sources 
disfavors the point source origin.  

Most of the soft diffuse sources have dynamical time scales of $\sim$10$^5$ yr, which  
corresponds to the last epoch of the high star formation activity of $\sim$10$^5$--10$^7$ yr ago  \citep{Yusef2009}. 
The abundances of Si and S of NE and NW and most of other soft diffuse sources are larger 
than CAs (typically $\sim$0.2 solar, e.g., \cite{Pandey2008}), but are typical  to those of a normal diffuse hot plasma. 
The horn-like sources NE and NW may be made by either  super-wind from multiple supernovae,
high activity of stellar wind from many high-mass stars in the GC region $\sim$10$^5$ yr ago \citep{Ponti2015},
or the past flares of Sgr A$^\star$ (e.g., \cite{Koyama2018}).  
The origin of a power-law component with the Fe I K$\alpha$ line (6.4 keV) would also be the same activities near the GC. 

%The study of the large scale LTP enhancements in the GCXE and the spectral information, in particular the metal abundances, are largely limited at this moment. We, therefore, encourage future search and study of theLTP clump candidates. The instruments of SXI and SXS to be onboard XARM may play crucial roles for the study of the origin of the LTP. 

\section*{Acknowledgement}

The authors are grateful to all members of the Suzaku team. 
KKN is supported by Research Fellowships of JSPS for Young Scientists.
This work was supported by the Japan Society for the Promotion of Science (JSPS) 
KAKENHI Grant Numbers JP24540232 (SY) and JP16J00548 (KKN).

%%%
% See the manual for the detail.
%%%


\begin{thebibliography}{}
% Journals(e.g. A\&A,ApJ,AJ,NMRAS,PASP ...)
% Authors, Year, Journal, Vol#, Page#
% Journal Title Abbreviation >> http://www.asj.or.jp/pasj/Jabb.html
%\bibitem[Ak et al.(2008)]{Ak2008}
%   Ak, T., Bilir, S., Ak, S., \& Eker, Z. 2008, New Astron., 13, 133
\bibitem[Anders \& Grevesse(1989)]{Anders1989}
   Anders, E., \& Grevesse, N. 1989, Geochim. Cosmochim. Acta, 53, 197
%\bibitem[Audard et al.(2003)]{Audard2003}
%   Audard, M., G\"udel, M., Sres, A., Raassen, A. J. J., \& Mewe, R. 2003, \aap, 398, 1137
%\bibitem[Balucinska-Church \& McCammon(1992)]{bcmc1992} 
%   Balucinska-Church, M., \& McCammon, D. 1992, \apj, 400, 699
%\bibitem[Bernardini et al.(2012)]{Bernardini2012}
%   Bernardini, F., de Martino, D., Falanga, M., Mukai, K., Matt, G., Bonnet-Bidaud, J. -M., Masetti, N., \& Mouchet, M.
%   2012, \aap, 542, A22
%\bibitem[Bland-Hawthorn \& Cohen(2003)]{Bland2003}
%   Bland-Hawthorn, J., \& Cohen, M. 2003, \apj, 582, 246
%\bibitem[Dutra \& Bica(2000)]{Dutra2000}
%   Dutra, C. M., \& Bica, E. 2000, \aap, 359, L9
%\bibitem[Eze(2015)]{Eze2015}
%   Eze, R. N. C. 2015, Mem. S. A. It., 86, 96
%\bibitem[Ezuka \& Ishida(1999)]{Ezuka1999}
%   Ezuka, H., \& Ishida, M.\ 1999, \apjs, 120, 277
\bibitem[Gillessen et al.(2009)]{Gillessen2009}
   Gillessen, S., Eisenhauser, F., Trippe, S., Alexander, T., Genzel, R., Martins, F., \& Ott, T. 2009, \apj, 692, 1075
\bibitem[G\"udel(2004)]{Gudel2004}
   G\"udel, M. 2004, A\&A review, 12, 71
\bibitem[Heard \& Warwick(2013)]{Heard2013}
   Heard, V., \& Warwick, R. S. 2013, \mnras, 434, 1339
%\bibitem[Hellier \& Mukai(2004)]{Hellier2004}
%   Hellier, C., \& Mukai, K. 2004, \mnras, 352, 1037
%\bibitem[Hellier et al.(1998)]{Hellier1998}
%   Hellier, C., Mukai, K., \& Osborne, J. P. 1998, \mnras, 297, 526
\bibitem[Hong(2012)]{Hong2012}
   Hong, J. 2012, \mnras, 427, 1633
\bibitem[Ishisaki et al.(2007)]{Ishisaki2007}
   Ishisaki, Y., et al. 2007, \pasj, 59, 113
%\bibitem[Johnson et al.(2009)]{Johnson2009}
%   Johnson, S. P., Dong, H., \& Wang, Q. D. 2009, \mnras, 399, 1429
%\bibitem[Kaastra \& Mewe(1993)]{Kaastra1993}
%   Kaastra, J. S., \& Mewe, R. 1993, \aaps, 97, 443
%\bibitem[Koyama et al.(1989)]{Koyama1989}
%   Koyama, K., Awaki, H., Kunieda, H., Takano, S., Tawara, Y., Yamauchi, S., 
%   Hatsukade, I., \& Nagase, F. 1989, \nat, 339, 603
%\bibitem[Koyama et al.(1996)]{Koyama1996}
%   Koyama, K., Maeda, Y., Sonobe, T., Takeshima, T., Tanaka, Y., \& Yamauchi, S.
%   1996, \pasj, 48, 249
\bibitem[Koyama et al.(2007)]{Koyama2007a}
   Koyama, K., et al.\ 2007, \pasj, 59, S23
%\bibitem[Koyama et al.(2007b)]{Koyama2007b}
%   Koyama, K., et al. 2007b, \pasj, 59, S245
\bibitem[Koyama(2018)]{Koyama2018}
   Koyama, K. 2018, \pasj, 70, 1
%\bibitem[Kushino et al.(2002)]{Kushino2002}
%   Kushino, A., Ishisaki, Y., Morita, U., Yamasaki, N. Y., Ishida, M., Ohashi, T., \& Ueda, Y.
%   2002, \pasj, 54, 327
\bibitem[Mitsuda et al.(2007)]{Mitsuda2007}
   Mitsuda, K., et al.\ 2007, \pasj, 59, S1
%\bibitem[Lu et al.(2003)]{Lu2003}
%   Lu, F. J., Wang, Q. D., \& Lang, C. C. 2003, \aj, 126, 319
\bibitem[Mori et al.(2008)]{Mori2008}
   Mori, H., Tsuru, T. G., Hyodo, Y., Koyama, K., \& Senda, A. 2008, \pasj, 60, 183
\bibitem[Mori et al.(2009)]{Mori2009}
   Mori, H., Hyodo, Y., Tsuru, T. G., Nobukawa, M., \& Koyama, K.. 2009, \pasj, 61, 687
\bibitem[Muno et al.(2003)]{Muno2003}
   Muno, M. P., et al. 2003, \apj, 589, 225
%\bibitem[Muno et al.(2004)]{Muno2004}
%   Muno, M. P., et al. 2004, \apj, 613, 326
\bibitem[Muno et al.(2009)]{Muno2009}
   Muno, M. P., et al. 2009, \apjs, 181, 110
%\bibitem[Nakajima et al.(2008)]{Nakajima2008}
%   Nakajima, H., et al. 2008, \pasj, 60, S1
\bibitem[Nakashima et al.(2013)]{Nakashima2013}
   Nakashima, S., Nobukawa, M.,  Uchida, H., Tanaka, T.,  Tsuru, T. G., Koyama, K., Murakami, H., \& Uchiyama, H.
   2013, \apj, 773, 20
\bibitem[Nakashima et al.(2014)]{Nakashima2014}
   Nakashima, S., Nobukawa, M., Uchida, H., Tanaka, T., Tsuru, T. G., Koyama, K., Uchiyama, H., \& Murakami, H. 
   2014, Proceedings of the International Astronomical Union, Vol. 303, 349
%\bibitem[Nobukawa et al.(2010)]{Nobukawa2010}
%   Nobukawa, M., Koyama, K., Tsuru, T. G., Ryu, S. G., \& Tatischeff, V. 2010, \pasj, 62, 423
\bibitem[Nobukawa et al.(2016)]{Nobukawa2016}
   Nobukawa, M., Uchiyama, H., Nobukawa, K. K., Yamauchi, S., \& Koyama, K. 2016, \apj, 833, 268
\bibitem[Pandey \& Singh(2008)]{Pandey2008}
   Pandey, J. C., \& Singh, K. P. 2008, \mnras, 387, 1627
%\bibitem[Pandey \& Singh(2012)]{Pandey2012}
%   Pandey, J. C., \& Singh, K. P. 2012, \mnras, 419, 1219
\bibitem[Park et al.(2004)]{Park2004}
   Park, S., Muno, M. P., Baganoff, F. K., Maeda, Y., Morris, M., Howard, C., 
   Bautz, M. W., \& Garmire, G. P. 2004, \apj, 603, 548
%\bibitem[Ponti et al.(2010)]{Ponti2010}
%   Ponti, G., Terrier, R., Goldwurm, A., Belanger, G., \& Trap, G. 2010, \apj, 714, 732
\bibitem[Ponti et al.(2015)]{Ponti2015}
   Ponti, G., et al. 2015, \mnras, 453, 172
\bibitem[Reid(1993)]{Reid1993}
   Reid, M. J. 1993, \araa, 31, 345
\bibitem[Revnivtsev et al.(2007)]{Revnivtsev2007}
   Revnivtsev, M., Vikhlinin, A., \& Sazonov, S. 2007, \aap, 473, 857
\bibitem[Revnivtsev et al.(2009)]{Revnivtsev2009}
   Revnivtsev, M., Sazonov, S., Churazov, E., Forman, W., Vikhlinin A., \& Sunyaev, R.
   2009, \nat, 458, 1142
\bibitem[Sakano(2000)]{Sakano2000}
   Sakano, M. 2000, Ph. D. thesis, Kyoto University
\bibitem[Sakano et al.(2002)]{Sakano2002}
   Sakano, M., Koyama, K., Murakami, H., Maeda, Y., \& Yamauchi, S. 2002, \apjs, 138, 19
\bibitem[Sazonov et al.(2006)]{Sazonov2006}
   Sazonov, S., Revnivtsev, M., Gilfanov, M., Churazov, E., \&  Sunyaev, R. 2006, \aap, 450, 117
\bibitem[Serlemitsos et al.(2007)]{Serlemitsos2007}
   Serlemitsos, P., et al.\ 2007, \pasj, 59, S9
%\bibitem[Sofue(1989)]{Sofue1989}
%   Sofue, Y. 1989, The Center of the Galaxy (Dordrecht: Kluwer), 213
%\bibitem[Su et al.(2010)]{Su2010}
%   Su, M., Slatyer, T. T., \& Finkbeiner, D. P. 2010, \apj, 724, 1044
%\bibitem[Sunyaev et al.(1993)]{Sunyaev1993}
%   Sunyaev, R., Markevitch, M., \& Pavlinsky, M. 1993, \apj, 407, 606
\bibitem[Tawa et al.(2008)]{Tawa2008}
   Tawa, N., et al. 2008, \pasj, 60, S11
\bibitem[Tsuru et al.(2009)]{Tsuru2009}
   Tsuru, T. G., Nobukawa, M., Nakajima, H., Matsumoto, H., Koyama, K., Yamauchi, S. 2009, \pasj, 61, S219
\bibitem[Uchiyama et al.(2009)]{Uchiyama2009}
   Uchiyama, H., et al. 2009, \pasj, 61, S9
\bibitem[Uchiyama et al.(2013)]{Uchiyama2013}
   Uchiyama, H., Nobukawa, M., Tsuru, T. G., \& Koyama, K.
   2013, \pasj, 65, 19
\bibitem[Wang et al.(2002)]{Wang2002}
   Wang, Q. D., Gotthelf, E. V., \& Lang, C. C. 2002, \nat, 415, 148
\bibitem[Warwick(2014)]{Warwick2014}
   Warwick, R. S. 2014, \mnras, 445, 66
%\bibitem[Xu et al.(2016)]{Xu2016}
%   Xu, X.-J., Wang, Q. D., \& Li, X.-D. 2016, \apj, 818, 136
%\bibitem[Yamauchi et al.(1990)]{Yamauchi1990}
%   Yamauchi, S. Kawada, M., Koyama, K., Kunieda, H., Tawara, Y., \& Hatsukade, I. 
%   1990, \apj, 365, 532
%\bibitem[Yamauchi et al.(2014)]{Yamauchi2014}
%   Yamauchi, S., Shimizu, M., Nakashima, S., Nobukawa, M., Tsuru, T. G., Koyama, K.  2014, \pasj, 66, 125
\bibitem[Yamauchi et al.(2016)]{Yamauchi2016}
   Yamauchi, S., Nobukawa, K. K., Nobukawa, M., Uchiyama, H., \& Koyama, K. 2016, \pasj, 68, 59
%\bibitem[Yasui et al.(2015)]{Yasui2015}
%   Yasui, K., et al. 2015, \pasj, 67, 123
\bibitem[Yusef-Zadeh et al.(2009)]{Yusef2009}
   Yusef-Zadeh, F., et al. 2009, \apj, 702, 178
\end{thebibliography}
\end{document}